# Influence of Nonmetal Dopants on Charge Separation of Graphitic Carbon Nitride by Time-Dependent Density Functional Theory

Tzu-Jen Lin*[a,b]

* Corresponding author: tzujenlin999@gmail.com

[a] Department of Chemical Engineering, Chung Yuan Christian University, 200 Chung Pei Road, Chung Li District, Taoyuan City, 32023, Taiwan
[b] Luh Hwa Research Center for Circular Economy, 200 Chung Pei Road, Chung Li District, Taoyuan City, 32023, Taiwan

## Abstract

Photocatalysts are crucial materials for green energy production and environmental purification. Non-metal doped graphitic carbon nitride (g-$C_3N_4$) has attracted much attention in recent years because of low-cost and desired photocatalytic performance characteristics such as high charge separation efficiency and broad visible light absorption. In this study, we used time-dependent density functional theory and wave function analysis to evaluate the charge separation characteristics of phosphorus-, oxygen- and sulfur-doped g-$C_3N_4$ upon photon excitation based on electron-hole pair distances, electron-hole pair overlaps, and charge transfer amounts. The lowest unoccupied molecular orbital of doped heptazine rings was shifted downward to facilitate electron transfer from undoped to doped heptazine rings upon photon excitation. Generally, the phosphorus dopant yielded relatively high charge transfer, high electron-hole pair separations, and low electron-hole overlap compared with the oxygen and sulfur dopants. In a low concentration, the sulfur dopant had the similar performance as the phosphorus dopant did. Furthermore, the phosphorus dopant attracted photon excited electrons, whereas the oxygen and sulfur dopants contributed to electron-hole pair separation. The dopants concentrated in one heptazine ring exhibited high charge separation performance than those dopants distributed among multiple heptazine rings in g-$C_3N_4$; the simulation results suggested that doping configurations are more crucial than doping concentrations with respect to charge separation efficiency in nonmetal-doped g-$C_3N_4$.

1. Introduction

The elimination of fossil fuels is a crucial goal of many contemporary societies because these fuels are nonrenewable, and their continued use could lead to energy crisis. In addition, the consumption of fossil fuels contributes to severe environmental problems such as air pollution and the greenhouse effect. Consequently, solar energy, which is both economical and inexhaustible, is often considered as a favorable alternative energy source. Photocatalysts are essential materials to convert solar energy into useable chemical energy. In 1972, Fujishima and Honda split water under ultraviolet (UV) light by using a $TiO_2$ electrode.[1] Upon photon excitation, the photon-generated electrons and holes reduced hydrogen ion into hydrogen molecules and oxidized water respectively. Since then, semiconductor photocatalysts such as ZnO, $Fe_2O_3$, CdS, and ZnS, have attracted widespread attention.[2-5] Photocatalysts can also be used for environmental purification. The aforementioned photon-generated holes and electrons form hydroxyl and superoxide ions, respectively, to degrade pollutants into water and carbon dioxide. The band gaps of traditional semiconductor photocatalysts are all approximately 3.2 eV; thus, these photocatalysts use only UV light, which accounts for 4% of solar energy.[6-9] Although $Cu_2O$ and $\alpha$-$Fe_2O_3$ can efficiently use up to 42% of solar energy, they have low stability and activity. Furthermore, the limited amounts and toxicity of metals make semiconductor photocatalysts expensive and not environmentally friendly.

In 2009, Wang and coworkers were first researchers to report that graphitic carbon nitride (g-$C_3N_4$) can absorb visible light (2.7 eV) to generate hydrogen.[10] Since then, g-$C_3N_4$ nanomaterials have been considered as promising metal-free photocatalysts. g-$C_3N_4$ is the most stable carbon nitride allotropes under ambient condition. It is composed of heptazine units connected by amino groups, and contains earth-abundant elements for example, carbon, nitrogen, and hydrogen;[11, 12] consequently, it is low-cost, and has potentials to be commercialized. In addition, g-$C_3N_4$ demonstrates high chemical and thermal stability owing to its carbon-nitrogen conjugated covalent bond networks and interlayer van der Waals interactions. However, pristine g-$C_3N_4$ cannot reach the desired levels of photocatalytic performance because of its low absorption ability at long wavelengths, small surface area, and rapid recombination of photon-generated electron-hole pairs. Therefore, many modifications have been proposed to enhance its photocatalytic performances; these modifications include change to the microstructures of g-$C_3N_4$ and combinations of the material with other semiconductive materials.[13, 14]

Doping metals and nonmetals into photocatalysts are another ways to tune their photocatalytic performance. Carbon and nitrogen atoms in the heptazine rings of g-$C_3N_4$ are possible substitution sites for nonmetal dopants. Therefore, many attempts to employ nonmetal-doped g-$C_3N_4$ to create pure nonmetal photocatalysts have been made. Phosphorus,[15-18] oxygen,[19, 20] and sulfur[21, 22] are the most commonly used dopants for the production of nonmetal-doped g-$C_3N_4$. The key to creating high-performance photocatalysts is the suppression of electron-hole pair recombination upon photon excitation. Experimental studies have revealed that phosphorus, oxygen, and sulfur increase the charge separation efficiency through the photoluminescence spectrum. However, some questions remain, such as which dopants have the highest charge separation efficiencies, and what the mechanism of the aforementioned increase is. In experimental research, quantifying the charge separation performance and mechanisms of dopants are difficult. Cheng and Zhang demonstrated that oxygen-doped g-$C_3N_4$ had excellent charge separation performance,[19] Wang and Yuan revealed that sulfur-doped g-$C_3N_4$ had higher charge separation performance than phosphorus-doped g-$C_3N_4$ did,[23] and Hu showed that phosphorus-doped g-$C_3N_4$ had higher charge separation performance than sulfur- and oxygen-doped g-$C_3N_4$ did.[16] *Ab initio* methodologies are suitable for studying the electronic structure of nonmetal-doped g-$C_3N_4$, but previous focus points were the band gaps and density of state (DOS) analysis.[24-27] How dopants influence the charge separation performance of doped g-$C_3N_4$ upon photon excitation had yet to be studied in detail. In this study, we employed time-dependent density functional theory (TD-DFT) and wave function analysis to systematically analyze the charge separation performance of nonmetal-doped g-$C_3N_4$.

2. Model and Computational Method

Being composed of triazine or heptazine rings interconnected through tertiary amines, g-$C_3N_4$ can have a fully condensed or polymeric melon structure.[11, 12] Theoretical studies by Zhu and coworkers have revealed that the heptazine-ring-based structures are the most stable atomic arrangements of g-$C_3N_4$.[25] Therefore, we constructed g-$C_3N_4$ models based on heptazine rings. Hybrid functional are well-known for having higher performance than other types of functionals in excited state calculations.[28-31] Furthermore, hybrid functional are easier to implement in molecular rather than periodic systems. Thus, this study employed a molecular model to mimic heptazine-based g-$C_3N_4$. We constructed molecular models of a heptazine monomer (HM), a heptazine dimer (HD), and a heptazine trimer (HT); these models are depicted in Figs. 1 and 2.

The doping sites of phosphorus, oxygen, and sulfur were selected based on experimental studies. X-ray photoelectron spectroscopy (XPS) measurements revealed that the state of P-N coordination in the g-$C_3N_4$ framework;[15-18] however, the doping position of the phosphorus atom in g-$C_3N_4$ was unclear. Phosphorous atoms could replace either corner or bay carbon atoms in heptazine rings to form P-N bonds. Theoretical studies have indicated low defect formation energies when phosphorus replaced corner or bay carbon atoms.[25] Nevertheless, we expected that the corner position would not yield a stable structure. We substituted the corner carbon atom for the phosphorus atom in the HT model, and the optimized structure became severely twisted as shown in Fig. S1. Therefore, we did not consider this a reasonable structure, and instead we replaced the bay carbon atom with the phosphorus atom in the heptazine rings. XPS indicated that C-S and C-O bonds were formed after doping sulfur and oxygen atoms into g-$C_3N_4$, respectively. This finding suggested that sulfur or oxygen atoms can replace nitrogen atoms in g-$C_3N_4$. Based on the XPS results, we constructed molecular models of phosphorus-, oxygen-, and sulfur-doped HD and HT to mimic nonmetal-doped g-$C_3N_4$ (Figs. 1 and 2). Moreover, we considered multiple dopants to examine how doping concentrations and configurations influenced charge separation in low-lying excited states. Sulfur and oxygen atoms both possess one more valence electron than nitrogen atoms do. Therefore, when replacing a nitrogen atom in g-$C_3N_4$, a sulfur or oxygen atom must relinquish one electron to form stable covalent bonds. Therefore, sulfur and oxygen gave doped heptazine rings a positive charge. A similar scenario was observed for phosphorus-doped g-$C_3N_4$. The phosphorus atom possesses one more valence electron than the carbon atom does, and must relinquish one electron to form stable covalent bonds, leading to positively charged heptazine rings.

Benchmark studies have shown that the one-parameter PBE0[32] functional has high performance on n → π* and π → π* transitions, as well as ground state geometry in organic systems. Nevertheless, hybrid functional with less than 33% Hartree-Fock exact exchange may not completely describe charge transfer excitation.[28, 30, 31, 33, 34] Therefore, we optimized the geometries of all molecular models by applying the PBE0 functional with 6-31g(d) basis sets followed by linear response formalism TD-DFT calculations[35, 36] through high fraction Hartree-Fock exact exchange functional or long-range correction. The TD-DFT calculations were performed using the PBE01/3,[37] M06-2X,[38] and CAM-B3LYP[39] functionals with 6-311+g(2d,p) basis sets. After these calculations, we conducted wave function analysis to characterize the lowest excited states based on electron-hole pair distances, overlaps, and charge transfer levels between different heptazine units in g-$C_3N_4$. The electron-hole pair distances, Δr, were calculated based on the method developed by Adamo and Guido;[34] Δr values were based on the charge centroids of the orbitals involved in the excitations. Electron-hole pair overlaps, Λ, were calculated based on spatial overlaps between occupied and unoccupied orbitals involved in excitations.[40, 41] Moreover, natural transition orbitals (NTOs) were calculated to examine the spatial distributions of electron-hole pairs upon photon excitation.[42] Although DOS is not well defined in isolated systems, adequate broadening of discrete energy levels by Gaussian functions provides valuable information regarding orbital compositions.[43-45] All DFT and TD-DFT calculations were performed using Gaussian 16 software,[46] and all wave function analysis with respect to electronic excitations was conducted using Multiwfn 3.6.[47]

3. Results and Discussion

3.1 Heptazine monomer

The highest occupied molecular orbital (HOMO) and lowest unoccupied molecular orbital (LUMO) electronic structures of the HM are depicted in Fig. 3(a). The HOMO and LUMO were constituted by the out-of-phase $2P_z$ orbitals of nitrogen and carbon atoms, respectively. In addition, the HOMO and LUMO exhibited strong antibonding characteristics. The incorporation of phosphorus, oxygen, or sulfur altered the HOMO and LUMO electronic structures of HM. Fig. 3(b) shows that parts of the HOMO and LUMO shrank leading to fewer antibonding interactions in the phosphorus-doped HM (HM_P). In addition, some bonding interactions between carbon and nitrogen atoms were formed at the HOMO and LUMO of HM_P. No evident orbital size reductions of the HOMO and LUMO were observed in the oxygen-doped HM (HM_O) and sulfur-doped HM (HM_S) as shown in Fig 3(c) and (d). Nevertheless, we observed bonding interactions between carbon and nitrogen atoms at the HOMO and LUMO. The size reduction of the antibonding orbital and the formation of bonding interactions between carbon and nitrogen atoms led to an energy downshift in the HOMO and LUMO as listed in Table 1.

The S0 → S1 excitation energies of the doped HM were similar to or higher than that of the undoped HM. This finding was inconsistent with the experimental observation that non-metal dopants broadened the visible light absorption range (red shift) of g-$C_3N_4$.[15-22] Moreover, we observed no any charge transfer characteristics on the S0 → S1 excited states of the doped HM. These excited states demonstrated strong local excitation characteristics based on their low Δr values (< 2.0 Å).[34] The computed excitation energies and Δr values indicated that the doped HM models did not capture the features of nonmetal-doped g-$C_3N_4$ in the experiments; however, we observed that the nonmetal dopants altered the

electronic structures of the HOMO and LUMO in the HM

3.2 Heptazine Dimer

The S0→ S1 excitation energy decreased 0.2 to 0.45 eV after the incorporation of one nonmetal dopants into the HD by the M06-2X and CAM-B3LYP functionals. PBE01/3 exhibited a larger reduction in S0→ S1 excitation energy from 0.4 to 0.9 eV (Table 2). The phosphorus dopants exhibited the lowest S0→ S1 excitation energy among the three single-doped HD. The reduction in S0→S1 excitation energy was consistent with qualitative experimental observations. Pristine g-$C_3N_4$ demonstrated an absorption edge at 2.7 eV, and the absorption edges of phosphorus-, oxygen-, and sulfur-doped g-C3N4 exhibit red shift approximately 0.1 eV.[10, 15-22] The large red shift with respect to the experimental observations in the three single-doped HD by simulation was because the concentrations of the doped (mass percentages: HD, 7.1% for HD_P, 3.8% for HD_O, and 7.3% for HD_S) were larger than that of experiments (usually less than 5%). Fig. 4 shows the NTOs of the S0→ S1 excited states in the pristine HD and single-phosphorus-, single-oxygen-, and single-sulfur-doped heptazine dimers, HD_P, HD_O, and, HD_S, respectively. In the pristine HD, the excited electron and hole of the S0→ S1 transition located in one of the heptazine rings correspond with a high electron-hole pair overlap, Λ=0.52. By contrast, HD_P, HD_O, and HD_S demonstrated varying electron-hole pair patterns. The excited holes were concentrated on the nitrogen atoms in the undoped heptazine ring, and the excited electrons distributed both in the doped and undoped heptazine rings. These findings indicated that the S0→ S1 transitions were a mixture of local and charge transfer characteristics in HD_P, HD_O, and HD_S. Most of the excited electrons in HD_P were located near the doped phosphorus atom, whereas those in HD_O and HD_S were located at the bottom edges of the doped heptazine rings based on NTO contours. The differences in excited electron distribution indicated that phosphorus attracted excited electrons, whereas oxygen and sulfur contributed to electron transfers at carbon-nitrogen conjugated covalent bonds.

The S0→S1 excited states of HD_P, HD_O, and HD_S demonstrated charge transfer characteristics through their Λ, Δr, and charge transfer quantity; we observed that phosphorus had higher charge separation performance compared with oxygen and sulfur in the doped HD based on three findings. First, the Λ values were 0.25, 0.31, and 0.33 for HD_P, HD_O, and HD_S, respectively. The phosphorus dopant created lower electron-hole pair overlaps than the oxygen and sulfur dopants did. Second, the Δr values of HD_P, HD_O, and HD_S were 5 to 6 Å as listed in Table 2, and phosphorus dopant created larger electron-hole pair distances than the oxygen and sulfur dopants did. Third, the phosphorus dopant induced greater charge transfer quantity than oxygen and sulfur dopants did. The charge transfer quantity of S0→S1 excitation by the phosphorus dopants by PBE01/3 functional was 1.4 times higher than the oxygen and sulfur dopants, and 1.66 times higher than oxygen and sulfur dopants when using the M06-2X and CAM-B3LYP functionals.

When two dopants were evenly distributed on both rings of the pristine HD, HD_2P, HD_2O, and HD_2S, we observed different characteristics of S0→ S1 excited state of these double-doped HD. First, their excitation energies exhibited no evident red shift with respect to pristine HD (Table 2). Second, the charge transfer characteristics of the excited states were not present in HD_2P, HD_2O, and HD_2S. The Δr indices dropped to lower than 1 Å in HD_2P, whose Λ value was 0.54 which was close to that of pristine HD. In addition, the amounts of charge transfer during the S0→ S1 transition in HD_2P was 0. The Δr values of HD_2O and HD_2S obtained by PBE01/3 were considerably higher than the corresponding values obtained by the M06-2X and CAM-B3LYP functionals. This finding may be attributed to the self-interaction error within PBE0 functional. This error exaggerates charge transfer characteristics in low-lying excited states.[48, 49] The Δr values of HD_2O and HD_2S obtained by the M06-2X and CAM-B3LYP functionals ranged from 2 to 4 Å; these values seemed to indicate that few charge transfer characteristics were present during the S0→S1 transitions. However, these transitions in double-doped HDs were local excitations in each heptazine ring based on the excited electron and hole distributions in NTO contours as shown in Fig. 5. The Λ values of HD_2O and HD_2S were 0.51 and 0.42, respectively, indicating considerable overlaps between excited holes and electrons. In addition, the charge transfer amounts of HD_2O and HD_2S were considerably lower than those of HD_O and HD_S, respectively.

The diminished charge transfer characteristics of HD_2P, HD_2O, and HD_2S indicated that dopant concentration is not crucial for enhancing charge transfer upon photon excitations, and this finding was consistent with those of experimental observations. Fig. 6 details the DOS analyses of HD_P, HD_O, and HD_S. The HOMO and LUMO were constituted by undoped and doped heptazine rings, respectively, suggesting that the phosphorus, oxygen, and sulfur dopants shift the LUMO at the doped heptazine rings downward, thereby facilitating charge transfer from the HOMO at undoped heptazine rings to the LUMO at doped heptazine rings. In other words, the undoped and doped heptazine rings respectively became electron donors and acceptors after doping. This charge transfer scheme is presented in Fig. 6 (d). When each heptazine ring was doped, both the HOMO and LUMO shifted downward that, thereby rendering excitation across heptazine rings difficult. Therefore, HD_2P, HD_2O, and HD_2S all exhibited local rather than charge transfer excitation.

3.3 Heptazine Trimer

The S0→ S1 excitation energies of the single-doped HTs, HT_P, HT_O, and HT_S, decreased by less than 0.2 eV (0.2~0.4 eV through PBE1/3) with respect to undoped HT. (Table 3) The S0→ S1 excitation energies of the single-doped HTs exhibited red shift; however, the shift amounts were lower than those of the single-doped HDs because the dopant concentrations of the single-doped HTs (mass percentages: 4.9% for HT_P, 2.6% for HT_O, and 5.1% for HT_S) were considerably lower than those of single-doped HDs (mass percentages: 7.1% for HD_P, 3.8% for HD_O, and 7.3% for HD_S). The holes of the S0→ S1 transitions were distributed throughout the undoped heptazine rings in HT_P, HT_O, and HT_S. However, the excited electron distributions of the S0→ S1 excitations differed based on NTOs as presented in Fig. 7. As observed in HD_P, HD_O, and HD_S, phosphorus dopant attracted excited electrons, whereas the oxygen and sulfur dopants contributed to electron transfer at the edges of carbon-nitrogen conjugated covalent bonds. The features of the NTOs reflected their Δr values. For M06-2X, the Δr values of HT_P, HT_O, and HT_S were 3.79, 4.21, and 4.69 Å, respectively; CAM-B3LYP yielded similar results. In a low dopant concentration environment, the sulfur and oxygen dopants rendered longer electron-hole pair distances than phosphorus dopants did in the single-doped HT. Regarding the amounts of charge transfer, 0.376(0.350), 0.181(0.169), and 0.300(0.257) $e$ electrons were transferred from undoped to doped heptazine rings in HT_P, HT_O, and HT_S, respectively by M06-2X. (The values in the parenthesis are obtained by CAM-B3LYP.) The phosphorus dopant induced greater charge transfer than the oxygen and sulfur dopants did, as observed in the single-doped HDs. The charge transfer amounts in the single-doped HTs were lower than those of single-doped HDs because of the low dopant concentrations by M06-2X and CAM-B3LYP. The charge transfer amounts of HT_P and HT_O were 70% of those of HD_P and HD_O, respectively, and the corresponding amount of HT_S was not vastly different from that of HD_S. Based on the observed charge transfer amounts and electron-hole pair distances, we suggest that the phosphorus and sulfur dopant had higher charge separation performance than oxygen dopants did in a low concentration environment. Furthermore, the simulation results suggested that the dopants had different functions with respect to charge transfer upon photon excitation. The phosphorus dopant attracted excited electrons, whereas the oxygen and sulfur dopants facilitated charge separations upon photon excitation. These findings may be related to a synergistic effect that enhanced the phocatalytic performance of phosphorus-oxygen- and phosphorus-sulfur-codoped-g-$C_3N_4$ in

experimental studies.[23, 50, 51]

In the double-doped HTs, namely HT_2P, HT_2O, and HT_2S, the Δr, Λ, and charge transfer amounts were not considerably altered by M06-2X or CAM-B3LYP. The charge transfer amount in HT_2P in fact decreased with respect to HT_P. As addressed in the preceding section, the doped heptazine rings became electron acceptors that facilitated charge transfer from the undoped to doped heptazine rings. When two rings of a HT were doped, the trimer consisted of one electron donor and two electron acceptors. The number of electron donors was lower than that of electron acceptors in HT_2P, HT_2O, and HT_2S, leading to lower charge transfer amounts in the corresponding single-doped HTs, namely HT_P, HT_O, and HT_S respectively. Based on the Δr, Λ, and charge transfer results of HT_2P, HT_2O, and HT_2S, charge separation performance was similar in all three double-doped HTs.

We tested hypothetical models where two dopants were doped into one ring of HTs, termed HT_2P_B, HT_2O_B, and HT_2S_B. We observed long electron-hole pair distances, low electron-hole pair overlaps, high charge transfer amounts, and red shift of excitation energies in the three models. The converged dopants had higher charge separation performance than dispersed dopants did by the M06-2X and CAM-B3LYP functionals. The PBE01/3 functional exhibited the opposite trends, namely that dispersed doping had higher charge transfer performance than the converged doping did. We believe that this outcome was a result of the self-interaction error within the PBE01/3 functional leading to exaggerated charge transfer excitation.[48, 49] Because the double-doped rings of the HTs became strong electron acceptors, the charge transfer amounts of HT_2P_B, HT_2O_B, and HT_2S_B were all over 0.7 $e$. HT_2P_B had longer electron-hole pair distances than HT_2O_B and HT_2S_B did, as observed in the doped-HDs. When dopants were dispersed throughout three heptazine rings in the HT, termed HT_3P, HT_3O, and HT_3S, we observed no charge transfer characteristics of the S0→ S1 excitation. These results were consistent with those of HD_2P, HD_2O, and HD_2S presented in previous sections, and demonstrated that doping configurations exert more of an influence than doping concentrations do on charge separation in nonmetal-doped HDs and HTs.

4. Conclusion

In this theoretical study, we used molecular models to examine how phosphorus, oxygen, and sulfur dopants influenced the charge transfer excitation in g-$C_3N_4$ through linear response TD-DFT and wave function analysis. We analyzed Δr values, Λ values, and charge transfer amounts to quantify the charge transfer performance of

phosphorus-, oxygen-, and sulfur-doped g-$C_3N_4$. The dopants altered the electronic structures of the frontier molecular orbitals to facilitate charge transfer upon photon excitation. The LUMO of doped heptazine rings was shifted downward to facilitate electron transfer from the HOMO of undoped heptazine rings to the aforementioned LUMO. In other words, the undoped and doped heptazine rings respectively became electron donors and acceptors after doping. In the single doped HDs, the phosphorus dopant demonstrated longer electron-hole pair distances, lower electron-hole pair overlap, and a higher charge transfer amounts than the oxygen and sulfur dopants did. In the single-doped HTs which mimicked low dopant concentration environment, the phosphorus dopant exhibited a higher charge transfer amount, whereas the oxygen and sulfur dopants exhibited longer electron-hole pair distances during S0→S1 transitions. These findings indicated that different dopants had different functions during charge transfer excitation in a low concentration. The phosphorus dopant attracted excited electrons, whereas the oxygen and sulfur dopants contributed to charge separation upon photon excitation. These results may be related to synergistic effect in phosphorus-sulfur- and phosphorus-oxygen-codoped-g-$C_3N_4$. Generally, the phosphorus dopant exhibited higher performance than the oxygen and sulfur dopants did with respect to charge separation upon photon excitations. In a low concentration, the sulfur dopant had the similar performance as the phosphorus dopant did. This simulation study also revealed that increasing the dopant concentration did not guarantee enhanced charge transfer upon photon excitation; this finding was consistent with experimental observations. Converged doping demonstrated considerably higher charge separation performance than dispersed doping did. When a dopant was concentrated in one ring of HDs or HTs, the doped ring became a strong electron acceptor, leading to a substantial reduction in excitation energy, longer electron-hole pair distances, higher charge transfer amounts, and lower electron-hole pair overlaps. By contrast, the charge transfer characteristics were markedly diminished when the dopants were dispersed, leading to local excitation. Finally, this study demonstrated that TD-DFT is suitable for studying doped g-$C_3N_4$ systems. This methodology could be a promising tool for future research on water splitting and reactive oxygen species upon photon excitation.

5. Acknowledgements

This research was supported by the Ministry of Science and Technology, Taiwan (107-2221-E-033 -030). The author is grateful for the generous allocation of

Table 1. Highest occupied molecular orbital and lowest unoccupied molecular orbital energies, S0→S1 excitation energies (Ex), and electron-hole pair distances (Δr) of pristine and nonmetal-doped heptazine monomer obtained by the PBE01/3, M06-2X and CAM-B3LYP functionals. The units of energy and distance are eV and Å, respectively.

|      | PBE01/3 | | | M06-2X | | | CAM-B3LYP | | |
|------|---------|---|---|--------|---|---|-----------|---|---|
|      | HOMO LUMO | Ex | Δr | HOMO LUMO | Ex | Δr | HOMO LUMO | Ex | Δr |
| HM   | -7.02 -0.95 | 4.22 | 0.06 | -7.74 -0.50 | 4.22 | 0.10 | -7.86 -0.25 | 4.22 | 0.10 |
| HM_P | -11.58 -5.74 | 4.27 | 1.43 | -12.22 -5.25 | 4.35 | 1.08 | -12.32 -4.98 | 4.35 | 1.11 |
| HM_O | -11.59 -5.44 | 4.71 | 1.83 | -12.36 -4.97 | 4.76 | 1.66 | -12.45 -4.72 | 4.74 | 1.68 |
| HM_S | -11.39 -5.61 | 4.40 | 1.60 | -12.13 -5.14 | 4.46 | 1.43 | -12.22 -4.89 | 4.46 | 1.48 |

Table 2. S0→S1 excitation energies (Ex), electron-hole pair distances (Δr), and charge transfer amounts between the heptazine rings (Δq) of pristine and nonmetal doped heptazine dimers obtained by the PBE01/3, M06-2X, and CAM-B3LYP functionals. The units of energy, distance, and charges are eV, Å, and $e$, respectively.

|  | PBE01/3 | | | M06-2X | | | CAM-B3LYP | | |
| --- | --- | --- | --- | --- | --- | --- | --- | --- | --- |
|  | Ex | Δr | Δq | Ex | Δr | Δq | Ex | Δr | Δq |
| HD | 3.91 | 0.34 | 0 | 3.96 | 0.37 | 0 | 3.97 | 0.37 | 0 |
| HD_P | 3.07 | 6.26 | 0.734 | 3.50 | 5.68 | 0.518 | 3.55 | 5.56 | 0.479 |
| HD_O | 3.42 | 5.93 | 0.523 | 3.71 | 5.17 | 0.306 | 3.73 | 5.07 | 0.287 |
| HD_S | 3.50 | 5.73 | 0.539 | 3.78 | 5.04 | 0.329 | 3.82 | 4.93 | 0.302 |
| HD_2P | 3.81 | 0.59 | 0 | 3.97 | 0.62 | 0 | 3.98 | 0.67 | 0 |
| HD_2O | 4.09 | 5.33 | 0.261 | 4.35 | 2.89 | 0.035 | 4.35 | 2.68 | 0.027 |
| HD_2S | 3.79 | 5.78 | 0.297 | 4.07 | 4.60 | 0.146 | 4.08 | 4.49 | 0.132 |

Table 3. S0→S1 excitation energies (Ex), electron-hole pair distances (Δr), and charge transfer amount from the undoped heptazine rings to the doped heptazine rings (Δq) of the pristine and nonmetal-doped heptazine dimers obtained by the PBE01/3, M06-2X and CAM-B3LYP functionals. The units of energy, distance, and charges are eV, Å, and $e$, respectively.

|  | PBE01/3 | | | M06-2X | | | CAM-B3LYP | | |
|---|---|---|---|---|---|---|---|---|---|
|  | Ex | Δr | Δq | Ex | Δr | Δq | Ex | Δr | Δq |
| HT | 3.50 | 1.34 | 0.020 | 3.59 | 1.39 | 0.018 | 3.58 | 1.34 | 0.012 |
| HT_P | 3.09 | 4.45 | 0.648 | 3.44 | 3.79 | 0.376 | 3.47 | 3.72 | 0.350 |
| HT_O | 3.29 | 4.51 | 0.398 | 3.48 | 4.21 | 0.181 | 3.48 | 4.12 | 0.169 |
| HT_S | 3.16 | 5.82 | 0.626 | 3.48 | 4.69 | 0.300 | 3.50 | 4.51 | 0.257 |
| HT_2P | 2.98 | 6.37 | 0.884 | 3.47 | 3.81 | 0.301 | 3.50 | 3.76 | 0.278 |
| HT_2O | 3.23 | 6.50 | 0.830 | 3.48 | 4.60 | 0.256 | 3.49 | 4.55 | 0.246 |
| HT_2S | 3.46 | 5.82 | 0.593 | 3.70 | 4.62 | 0.281 | 3.73 | 4.63 | 0.264 |
| HT_2P_B | 1.79 | 6.41 | 0.730 | 2.34 | 6.21 | 0.711 | 2.40 | 6.16 | 0.705 |
| HT_2O_B | 1.98 | 5.46 | 0.790 | 2.54 | 5.33 | 0.739 | 2.53 | 5.29 | 0.736 |
| HT_2S_B | 2.10 | 5.37 | 0.790 | 2.57 | 5.32 | 0.727 | 2.60 | 5.26 | 0.705 |
| HT_3P | 3.60 | 0.89 | 0.010 | 3.72 | 0.87 | 0.002 | 3.72 | 0.92 | 0.002 |
| HT_3O | 4.05 | 2.46 | 0.005 | 4.17 | 1.94 | 0.005 | 4.16 | 1.96 | 0.003 |
| HT_3S | 3.72 | 2.63 | 0.030 | 3.87 | 2.44 | 0.011 | 3.87 | 2.42 | 0.011 |

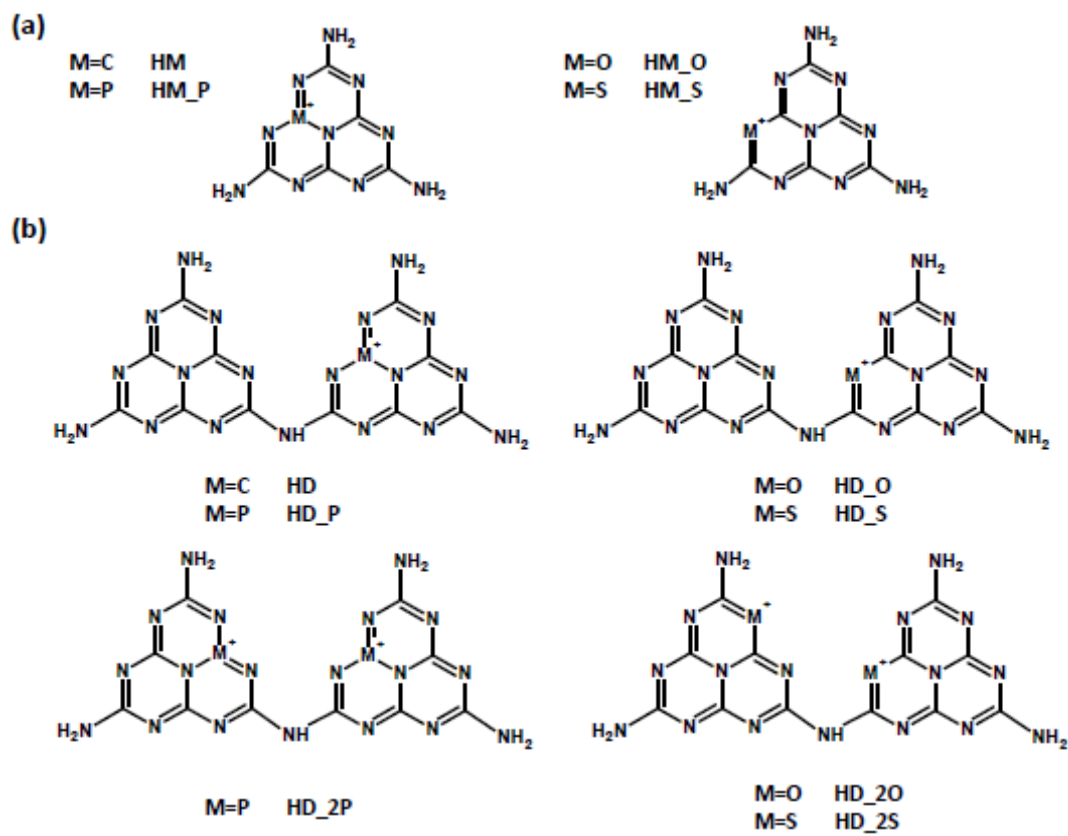

Fig. 1 Molecular models of undoped and non-metal doped (a) heptazine monomer and (b) heptazine dimer. The dopants were phosphorus (P), oxygen (O), and sulfur (S) atoms.

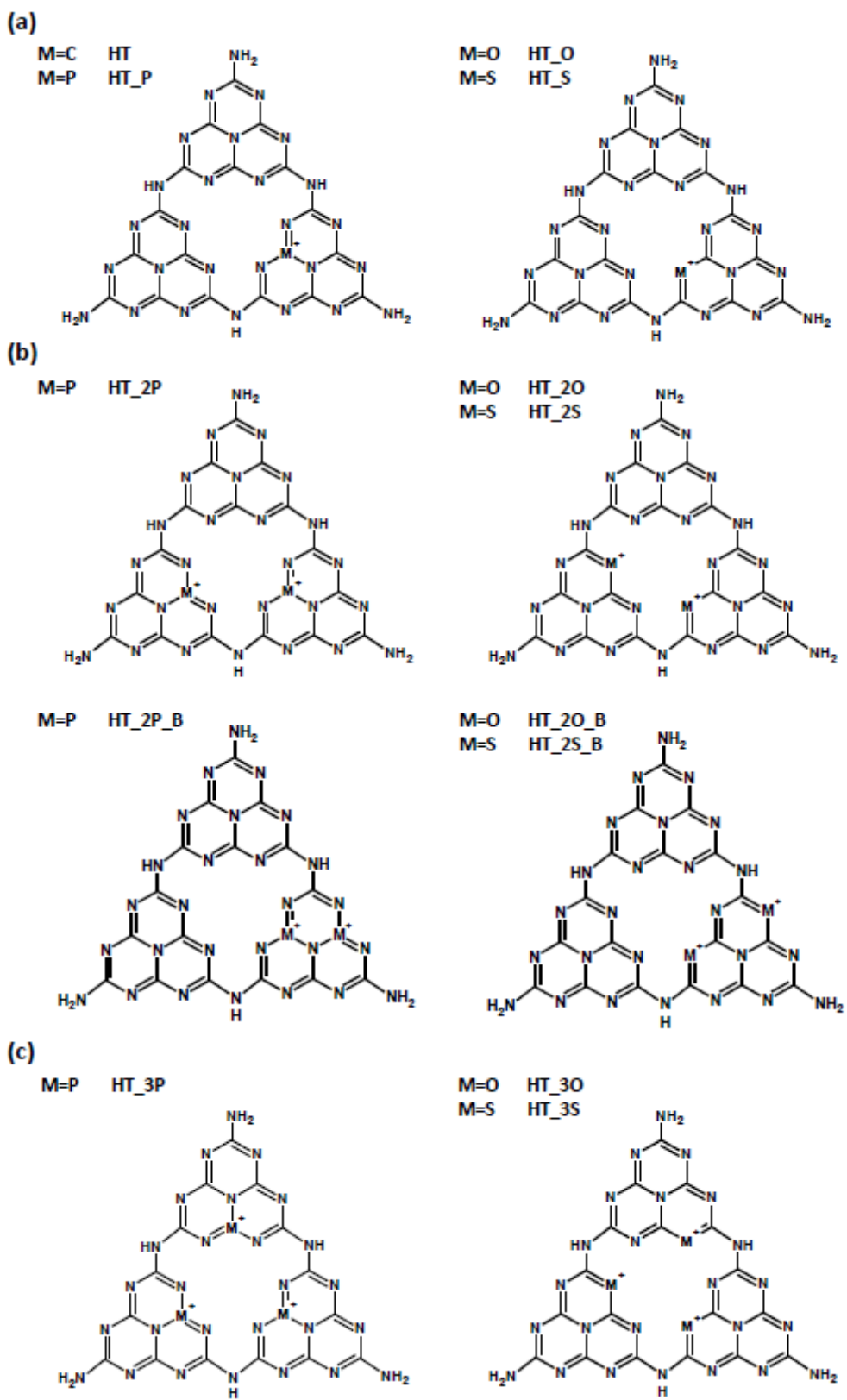

Fig. 2 Molecular models of (a) single doped, (b) double doped, and (c) triple doped heptazine trimers. The dopants were phosphorus (P), oxygen (O), and sulfur (S) atoms. Each dispersed double-doped heptazine dimer is marked with a "B."

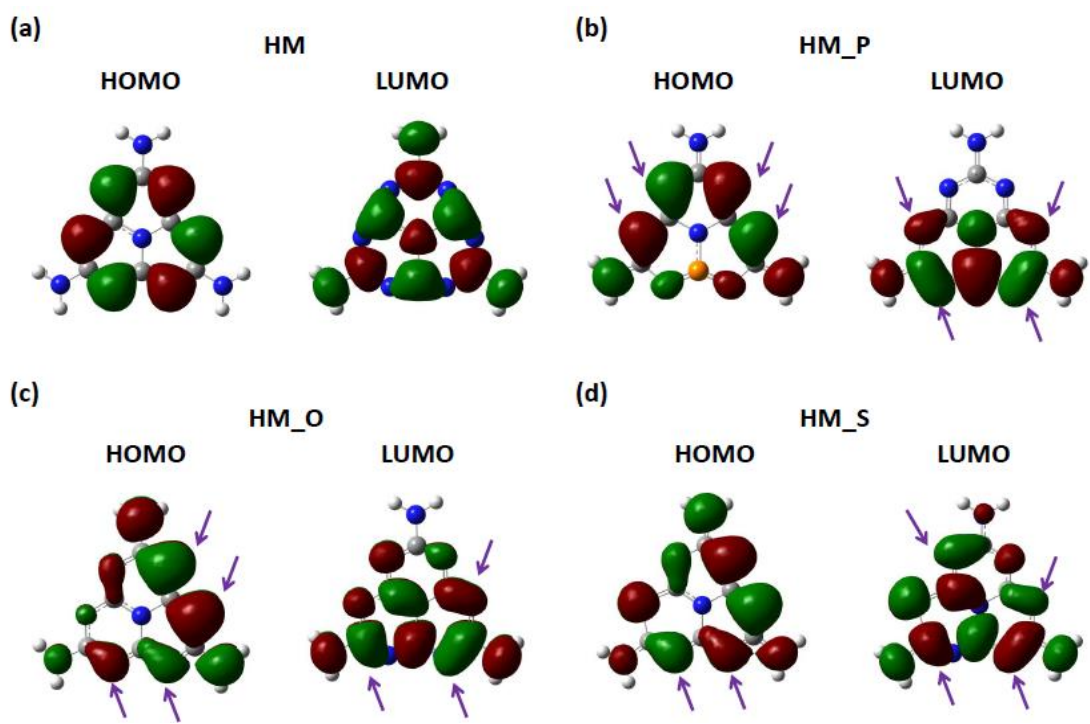

Fig.3 Highest occupied molecular orbital and lowest occupied molecular orbital electronic structures of (a) the pristine heptazine monomer, (b) the singe phosphorus doped heptazine monomer, (c) the single oxygen doped heptazine monomer, and (d) the single sulfur doped heptazine monomer. Isosurface value was 0.02 a.u. The arrows indicate points to bonding interactions between carbon and nitrogen atoms.

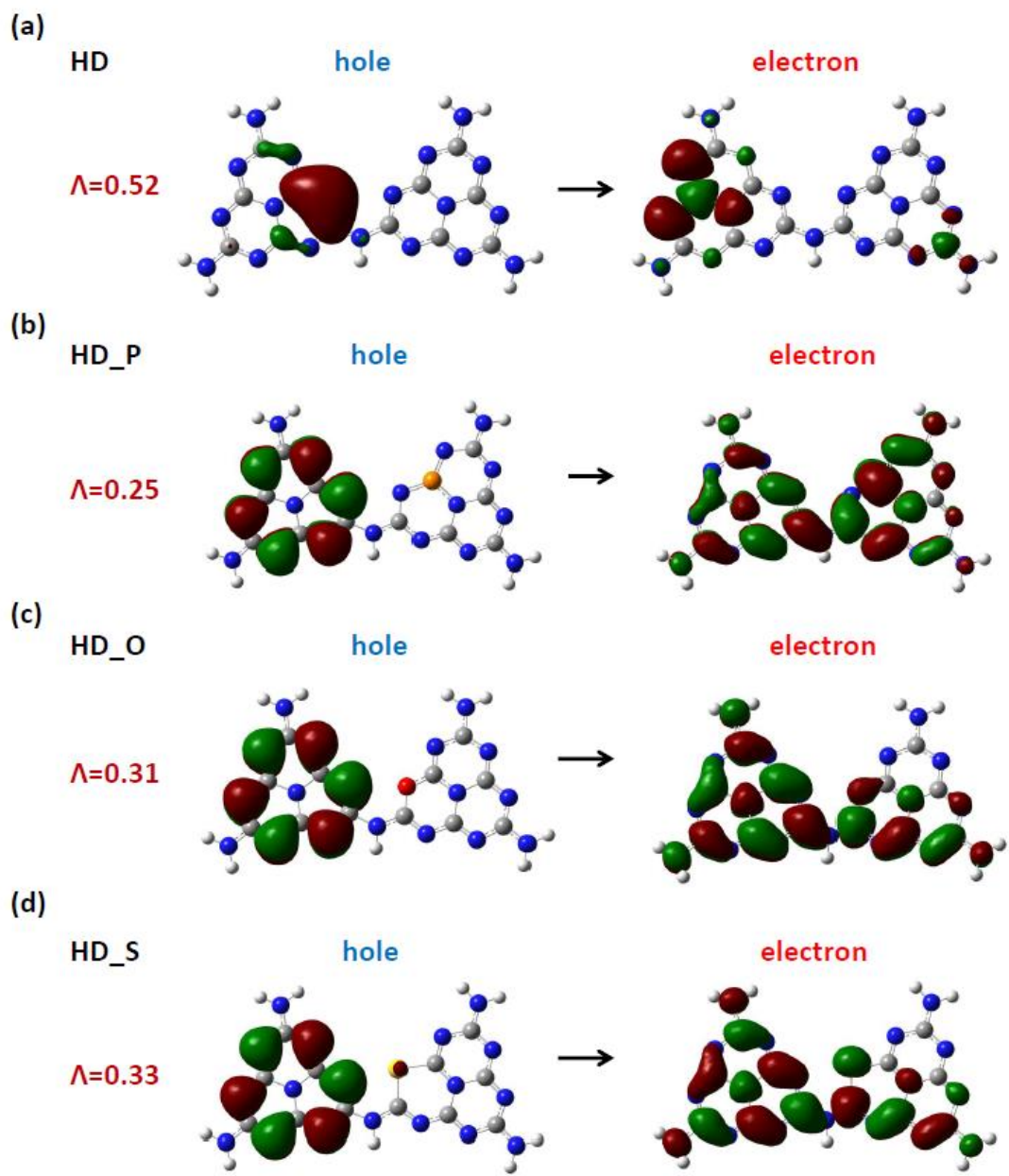

Fig. 4 Natural transition orbitals of (a) the pristine heptazine dimer, (b) the single-phosphorus-doped heptazine dimer, (c) the singe-oxygen-doped heptazine dimer, and (d) the single-sulfur-doped heptazine dimer. The natural transition orbitals were S0→S1 transitions, and the isosurface value was 0.02 a.u. Λ represents electron-hole pair overlap.

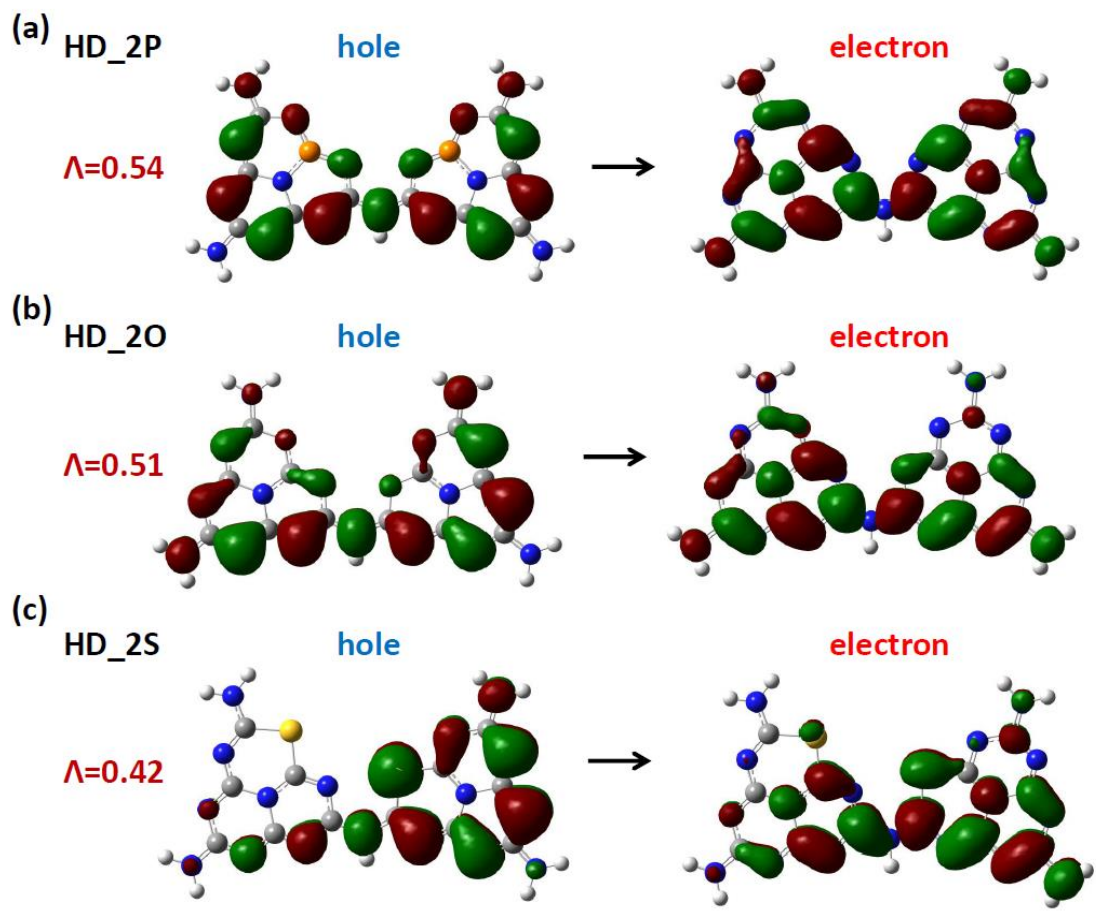

Fig. 5 Natural transition orbitals of (a) the double-phosphorus-doped heptazine dimer, (b) the double-oxygen-doped heptazine dimer, and (c) the double-sulfur-doped heptazine dimer. The natural transition orbitals were S0→S1 transitions, and the isosurface value was 0.02 a.u. Λ represents electron-hole pair overlap.

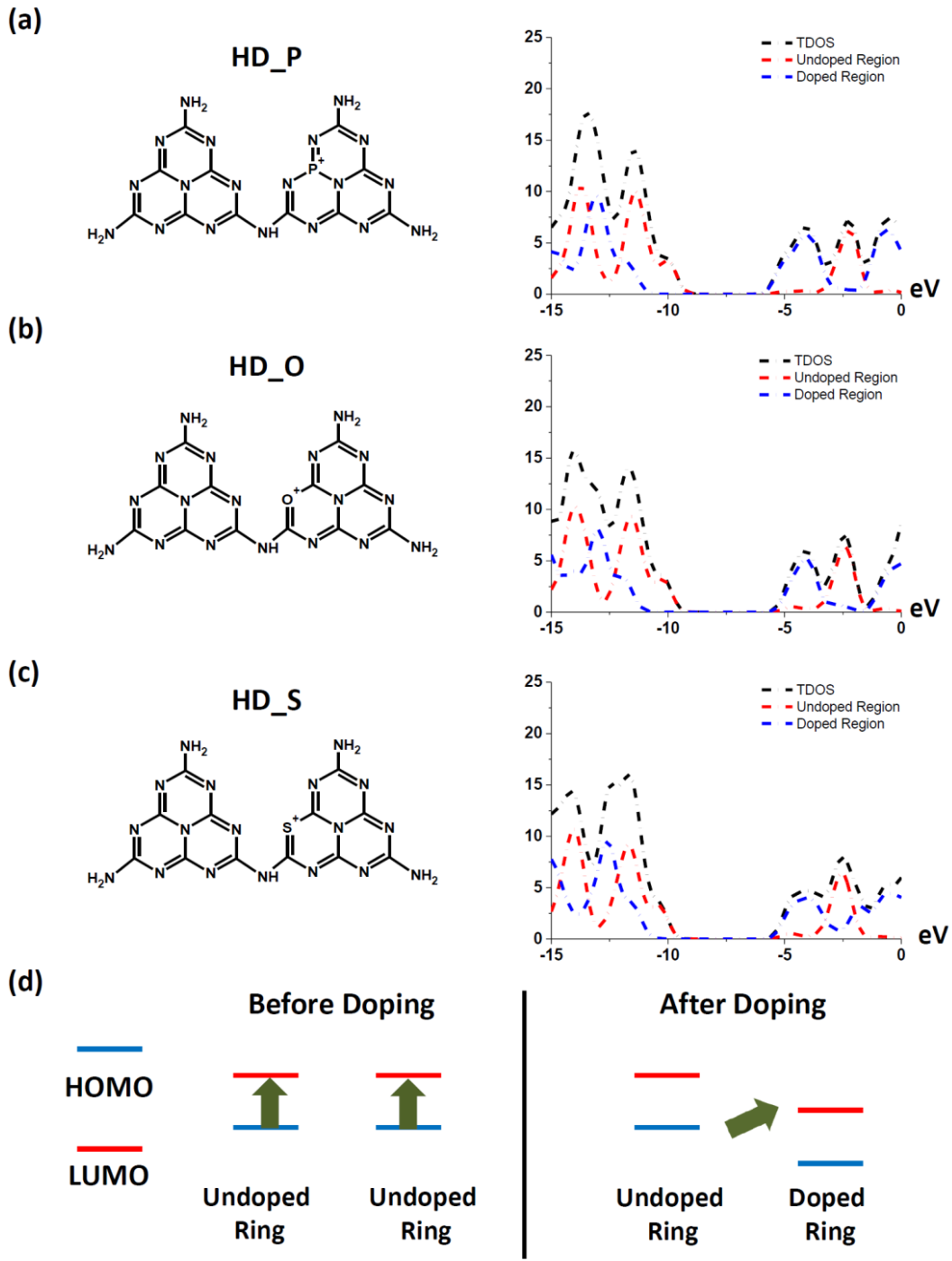

Fig. 6 Density of states for (a) the single-phosphorus-doped heptazine dimer, (b) the single-oxygen-doped heptazine dimer, and (c) the single-sulfur-doped heptazine dimer. (d) Schematic of electronic excitation in heptazine dimers before and after doping.

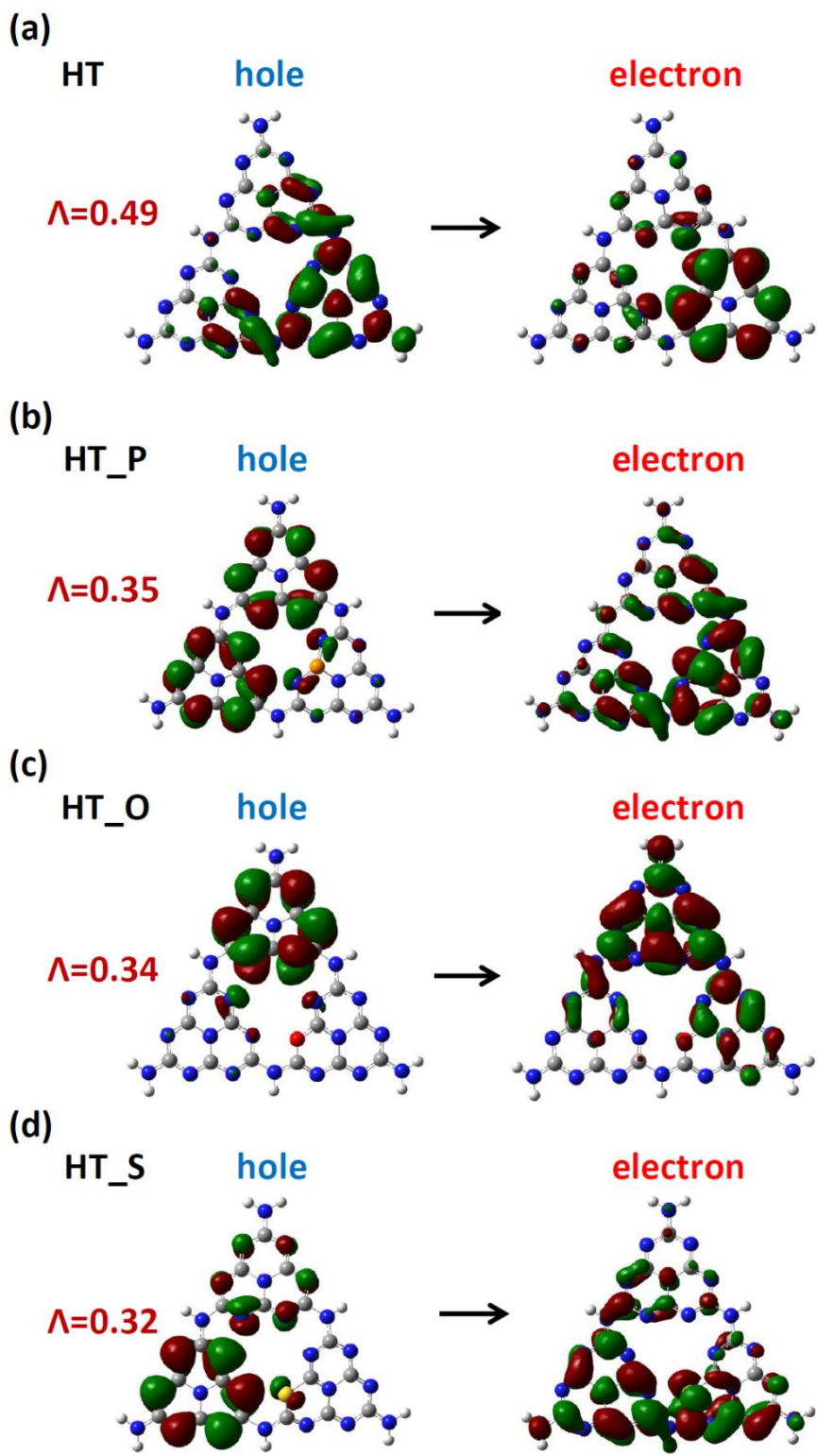

Fig. 7 Natural transition orbitals of (a) the pristine, (b) the single-phosphorus-doped, (c) single-oxygen-doped, and (d) the single-sulfur-doped heptazine trimer. The natural transition orbitals were S0→S1 transitions, and the isosurface value was 0.02 a.u. Λ represents electron-hole pair overlap.

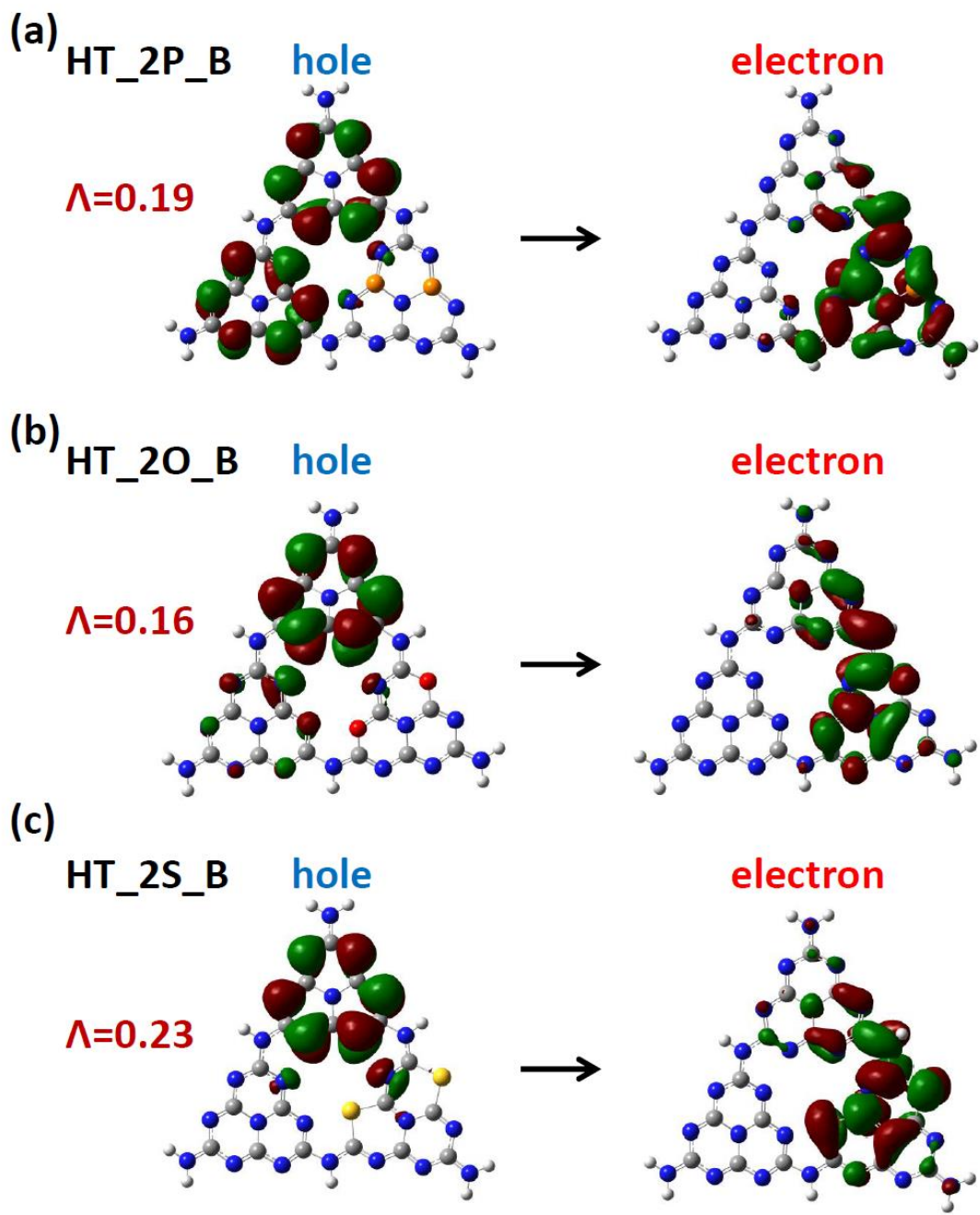

Fig. 8 Natural transition orbitals of the double-doped heptazine trimer where the dopant was concentrated in one heptazine ring. (a) Phosphorus dopant; (b) oxygen dopant; (c) sulfur dopant. The natural transition orbitals were S0→S1 transitions, and the isosurface value was 0.02 a.u. Λ represents electron-hole pair overlap.

# Supplementary Material
# for
# TD-DFT Study on the Influence of Nonmetal Dopants on Charge Separation of Graphitic Carbon Nitride

Tzu-Jen Lin*[a,b]

* Corresponding author: tzujenlin999@gmail.com

[a] Department of Chemical Engineering, Chung Yuan Christian University, 200 Chung Pei Road, Chung Li District, Taoyuan City, 32023, Taiwan
[b] Luh Hwa Research Center for Circular Economy, 200 Chung Pei Road, Chung Li District, Taoyuan City, 32023, Taiwan

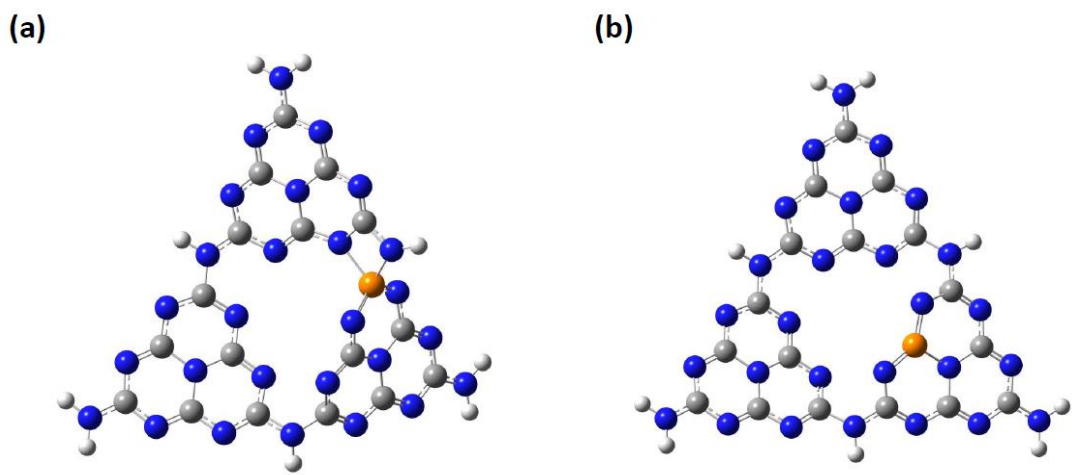

Fig. S1 The optimized ground state structure of the single-phosphorus-doped HT where the phosphorus atom was located at (a) the corner and (b) the bay positions.